\documentclass[preprint2]{aastex}

\shorttitle{Revisited sunspot activity in the 17-th century}
\shortauthors{Vaquero et al.}
\usepackage{natbib}

\begin{document}


\title{Revisited sunspot data: A new scenario for the onset of the Maunder minimum}


\author{Jos\'e M. Vaquero}
\affil{Departamento de F\'isica, Universidad de Extremadura, Spain}
\email{jvaquero@unex.es}

\author{M.C. Gallego}
\affil{Departamento de F\'isica, Universidad de Extremadura, Spain}
\email{maricruz@unex.es}

\author{Ilya G. Usoskin}
\affil{Sodankyl\"a Geophysical Observatory (Oulu unit), University of Oulu, Finland}
\email{ilya.usoskin@oulu.fi}

\and

\author{Gennady A. Kovaltsov}
\affil{Ioffe Physical-Technical Institute of RAS, 194021 St.Petersburg, Russia}
\email{gen.koval@mail.ru}




\begin{abstract}
Maunder Minimum forms an archetype for the Grand minima, and detailed knowledge of its
 temporal development has important consequences for the solar dynamo theory dealing with long-term solar activity evolution.
Here we reconsider the current paradigm of the Grand minimum general scenario by using newly recovered sunspot
 observations by G. Marcgraf and revising some earlier uncertain data for the period 1636--1642, i.e., one
 solar cycle before the beginning of the Maunder Minimum.
The new and revised data dramatically change the magnitude of the sunspot cycle just before the Maunder Minimum,
 from 60--70 down to about 20, implying a possibly gradual onset of the Minimum with reduced
 activity started two cycles before it.
This revised scenario of the Maunder Minimum changes, through the paradigm for Grand solar/stellar activity
 minima, the observational constraint on the solar/stellar dynamo theories focused on long-term studies and
 occurrence of Grand minima.
\end{abstract}





\section{Introduction}

The Sun is the only star whose features can be studied in great detail and on long time scale,
 forming a paradigm for a large population of 'sun-like' stars.
The Sun exhibits a great deal of magnetic variability generically called the solar activity,
 which is grossly dominated by the quasi-periodic $\approx$11-year cycle.
On top of that, secular variability is superposed ranging from very high activity during the late
 20-th century down to very quiet periods of Grand minima \citep{usoskin08,hathaway10}.
Generally, the 11-year cyclicity is understood in terms of solar dynamo \citep{charbonneau10}, while
 the secular/millenial variability still remains unclear.
A particular enigma for solar/stellar dynamo is the occurrence of Grand minima, that is not an
 intrinsic feature of the standard mean-field dynamo \citep{sokoloff04}.
Accordingly, the occurrence of Grand minima is often modelled by an {\it ad hoc} approach,
 including, e.g., stochastic or chaotic driven processes \citep[e.g.][]{moss08}.
Therefore, observational constraints on solar/stellar dynamo with respect to Grand minima are
 crucially important.
While the statistic of Grand minima/maxima of solar activity is studied to some extent using
 the cosmogenic isotope data \citep{eddy77,stuiver89,usoskin07,abreu08}, the variability of solar activity during
 and around a Grand minimum is not precisely known.
This is caused by the fact that presently we have more or less detailed information on only one
 Grand minimum - the Maunder minimum in the second half of 17-th century \citep{eddy76}, which serves
 as an archetype for Grand minima in general.
The present paradigm for a general scenario of a Grand minimum \citep[see][and references therein]{usoskin_mursula_03} is widely considered
 as a constraint for solar dynamo models.
Therefore, every piece of information on solar activity during that period is extremely valuable.

Temporal variability of solar activity around the Maunder minimum
 is usually studied using historical telescopic observations of sunspots since 1610 \citep{vaquero09}.
The period before the Maunder minimum (i.e., the first half of the 17-th century) was very uncertain
 in the original Wolf sunspot series, but the data were greatly improved after tremendous work
 of \citet{hoyt94} and \citet{hoyt98}, who collected large amount of historical sunspot records, forming the group sunspot number
 (GSN) series.
The solar activity variability before 1650 AD is shown in Fig.~\ref{Fig:t} according to the
 official GSN series (dotted curve) and reanalyzed with improved statistical techniques \citep[][-- grey curve]{usoskin03}.
The first observed solar cycle 1610--1618 was pretty high (maximum annual GSN values above 100), and this
 is quite reliable as it was covered by the direct data including
 sunspot drawings and counts \citep{hoyt98}.
The second cycle 1618--1630 was lower (maximum GSN values 30--40), but the quality of data is not good.
The next cycle 1635--1645 is marked as high in the GSN series, and this gave rise to the
 idea of an abrupt onset of the general Grand minimum scenario.

However, the GSN series is also not complete, and new data, which remained unnoticed by Wolf and his successors including
 Hoyt \& Schatten, are continuously recovered in various places, often outside major observatories \citep{arlt08,arlt09,casas06,vaquero04,vaquero05,vaquero07a}.
This leads to a revision of some parts of the GSN series and requires re-analysis of some results.
Here we reconsider the paradigm of Grand minimum general scenario by using newly recovered sunspot
 observations as well as revising some earlier uncertain data for the period 1636--1642, i.e., one
 solar cycle before the beginning of the Maunder minimum.

\section{Revised sunspot data for 1636--1642}

As discussed in great detail by \citet{vaquero07} sunspot data in 1610--1644 (before the Maunder Minimum) contains numerous
 gaps and uncertainties.
This problem is specially acute for the period from 1636 through 1642.
For that period the following data sets were available in the GSN series:
 (1) three years (1636, 1637 and 1641) without any solar observational record;
 (2) two years (1640 and 1642) with little amount of records by Horrox, Gassendi, Hevelius, Scheiner and Rheita
 (2 and 37, respectively);
 and (3) two years (1638 and 1639) with estimated data based on Crabtree data.
Here we use the GSN database (ftp://ftp.ngdc.noaa.gov/STP/SOLAR\_DATA/ SUNSPOT\_NUMBERS/GROUP\_SUNSPOT\_NUMBERS/),
 but critically revise it for 1636--1642.

First we add newly recovered sunspot data by G. Marcgraf which are not included
 into the GSN database.
Georg Marcgraf (1610--1648) was a German naturalist and astronomer \citep{north89}.
We have consulted his manuscripts of astronomical observations preserved in Leiden Regional Archive
(Collection Marcgraf LB7000/1) and
 in the National Library of Portugal (Mss. 6, n$^\circ$ 37).
We have recovered sunspot records for the year 1637, just the year with no records in the GSN database.
These records are preserved in the 'Collection Marcgraf' (Leiden Regional Archive, LB7000/1) in the
documents labelled as 22, 24a, 24b and 52 according to the numeration made by the historian of science
John North in the 1980s (North 1989).
Fig.~\ref{Fig:m} depicts an example of solar disc drawings by Marcgraf showing one sunspot from 9 to 12 June
1637 (document 24a).
Sunspot records by Marcgraf cover 17 days of 1637, that were included in the new database (see Table~\ref{Tab1}).

Another addition is correction of Horrox's records.
Jeremiah Horrocks or Horrox (1618--1641) was an English astronomer who predicted,
 observed and recorded the transit of Venus in 1639.
His sunspot records appeared in his 'Opera Posthuma' \citep{horroccii}.
Horrox used the Julian Calendar and, therefore, we have converted these dates to Gregorian Calendar \citep{vaquero07}.
Moreover, Horrox noted ''Maculae duas in Sole'' (two spots in the Sun) in 30 Oct -- 1 Nov 1638.
We have interpreted these two spots as one sunspot group to make this record compatible
 with the only group observed by Gassendi on the same dates.

Next we noted that the filling of the years 1638 and 1639 in GSN was done based on an unclear estimate
 rather than on real data.
\citet{hoyt98} wrote in their Bibliography appendix: ''According to a letter by Crabtree the average number
 of spot groups seen in 1638 and 1639 were 4-5 per day.
The database has Greenwich fill values to give 4-5 groups per day.
This substitution technique was used to simplify the analysis.
This is the only place in the entire database where we do this type of substitution.''
However, the number of the actually observed groups was smaller than the number of estimated groups for
 these years in the reconstruction \citep[see Figure 5 in][]{vaquero07}.
Therefore, we decided not to use this non-observed values for our estimations and eliminate Crabtree data.

We have also made some minor corrections to the database using the original sources.
We have eliminated one spurious observation by Gassendi on 1 Dec 1638, because this record does not appear
 in his astronomical observations \citep{gassendi}.
Moreover, we have incorporated one sunspot record by Horrox in 4 December 1639.
Curiously, the treatise "Venus in sole visa" [Venus in transit across the Sun] by Horrox was
 published by Johannes Hevelius at his own expense in 1662 \citep{hevelius}.
In this work, Horrox noted that he saw one small and common spot in the Sun on 4 Dec 1639 \citep[p. 115 of][]{hevelius}.
Finally, we have also changed the record by Rheita in 1642 using the original source \citep[pp. 242-243 of][]{rheita}.

All these additions and changes in the original Hoyt and Schatten database are listed in Table~\ref{Tab1} and
 summarized below:
\begin{enumerate}
\item Newly recovered sunspot records by G. Marcgraf are added;
\item The estimated (not observed) values from Crabtree's comments (1638--1639) are discarded;
\item Dates and numbers of sunspot groups are corrected for Horrox's observations;
\item A spurious observation by Gassendi on 1 December 1638 is eliminated;
\item One missing sunspot record by Horrox in 4 December 1639 is added.
\item The record by Rheita in 1642 is corrected.
\end{enumerate}

\begin{figure}
\begin{center}
\resizebox{8cm}{!}{\includegraphics{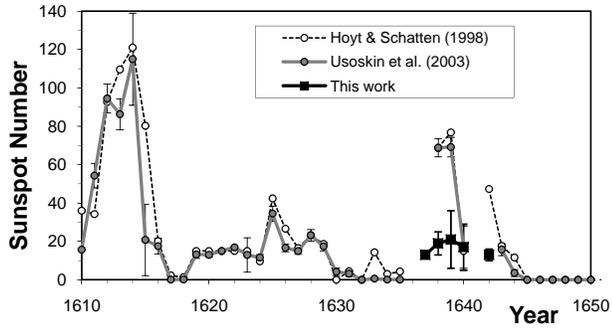}}
\end{center}
\caption{
Annual sunspot numbers in the first half of 17-th century.
Group sunspot numbers R$_{\rm g}$ \citep{hoyt98} are shown by the dotted line,
 weighted sunspot number, based on the same data set \citep{usoskin03} by grey line, and the weighted sunspot number
 R$_{\rm w}$ estimated in this work by black line.
\label{Fig:t}}
\end{figure}

\begin{figure}
\begin{center}
\resizebox{8cm}{!}{\includegraphics{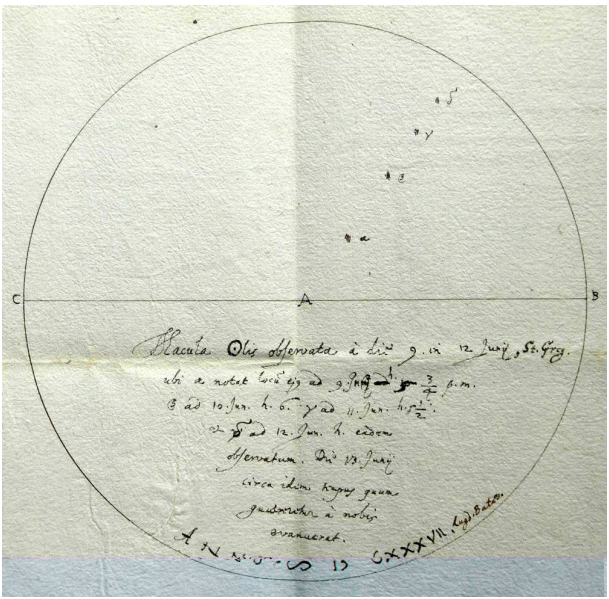}}
\end{center}
\caption{
An example of the solar disc drawing by Marcgraft depicting one sunspot from 9 to 12 June 1637.
\label{Fig:m}}
\end{figure}

Using the revised data presented in Table~\ref{Tab1}, we have evaluated the annual sunspot numbers for
 the period of 1637--1642, employing the statistical method proposed by \citet{usoskin03}.
The method is based on comparing the actual sparse data (sample population) to the daily sunspot
 data in 1850--1996 (reference population), assuming constancy of the statistical properties of sunspot activity.
For a given sample population of daily measurements within a month, months in the reference population
 are found that contain the same subset of daily values.
A statistical distribution of the corresponding monthly means is then built that allows to estimate the
 mean and uncertainty of monthly sunspot numbers, reconstructed from sparse daily observations
 \citep[see full details in][]{usoskin03}.
From monthly mean values, yearly values can be obtained in the same way.
The newly computed yearly sunspot numbers for 1637--1642 are given in Table~\ref{Tab2} and shown
 in Fig.~\ref{Fig:t} as thick solid line with error bars.

As one can see from Fig.~\ref{Fig:t}, the revised and updated data for 1637--1642 has essentially
 changed the profile of temporal variability of sunspot data before the Maunder minimum.
In particular, the last solar cycle before the minimum now appears quite modest, with the
 peak value of $20\pm 15$ compared to about 70 in the GSN series.
The new scenario, in accordance with the new database, implies low solar activity
 roughly two solar cycles before the beginning of the Maunder minimum.
This suggests that transition from the normal activity to the deep minimum was not as sudden as
 previously thought \citep{usoskin08}, and the descent of solar activity might have started as early
 as in the 1610s.

\section{Discussion and Conclusions}

Since the Maunder minimum forms an archetype for the Grand minima, detailed knowledge of its
 temporal development has important consequences for the solar dynamo theory dealing with long-term solar activity evolution.
The general dynamo theory \citep[see, e.g.][]{charbonneau10} cannot naturally reproduce occurrence
 of Grand minima and requires some prescribed changes in the dynamo parameters.
The present paradigm for the Maunder minimum \citep[e.g.][]{vitinsky86,sokoloff94,frick96,usoskin00,usoskin08}
 is that transition from the normal high activity to the deep minimum was sudden (within a few years) and
 without any apparent precursor, while the recovery to the normal activity level was gradual, taking several decades.
The abrupt onset of a Grand minima forms a strong constraint, as only few models with stochastically driving
 forces can 'naturally' produce such a feature \citep[e.g.][]{charbonneau01}, while others require special {\it ad hoc}
 assumptions.
Presently, several models can reproduce, with different approaches, the proposed scenario of a Grand minimum
 \citep[e.g.][]{charbonneau01,charbonneau04,usoskin_moss_09,karak10,passos11}.

On the other hand, many dynamo models succeed in predicting Grand minima to start gradually, through
 a continuous decrease of the activity level to the deep minimum, followed by a gradual recovery.
Just to mention a few, \citet{kueker99,weiss00,brooke02,bushby06,moss08}.
Thus, it is crucial to know the solar activity evolution before the Maunder Minimum with high confidence.

The revised sunspot records presented here imply that the scenario of the Maunder Minimum can
 correspond to a gradual onset, contrary to the earlier consideration, thus affecting observational constraints
 on the solar/stellar dynamo.
Unfortunately, cosmogenic isotope data can hardly resolve individual cycles \citep{steig96,usoskin09} to
 clearly answer this question, but a cautious statistical study of radiocarbon $^{14}$C data in tree rings
 for the Maunder and Sp\"orer solar minima indicate a possible lengthening and attenuation of solar cycles a
 few decades before the onset of a Grand minimum \citep{miyahara08}.

The major findings of this work can be summarized as follows.
\begin{enumerate}
\item
Using newly recovered sunspot records by Georg Marcgraf and carefully revised data for other
 observations (Table \ref{Tab1}), we provide a new sunspot number series (Table~\ref{Tab2}) for
 the period 1636--1642.
\item
The new data dramatically change the magnitude of the sunspot cycle just before the Maunder Minimum,
 from 60--70 down to about 20, implying a possibly gradual onset of the Minimum with reduced
 activity started two cycles before it.
\item
This revised scenario of the Maunder Minimum changes, through the paradigm for Grand solar/stellar activity
 minima, the observational constraint on the solar/stellar dynamo theories focused on long-term studies and
 occurrence of Grand minima.
\end{enumerate}

Thus, we have essentially revised the sunspot data prior to the Maunder Minimum leading to the
 revisited observational scenario of a Grand minimum of solar activity.
The present results are expected to impact development of models dealing with long-term solar/stellar
 activity evolution.



\acknowledgments
All the historical materials used in this work were consulted at the Regionaal Archief (Leiden, Netherlands),
 Biblioteca Nacional de Portugal (Lisbon, Portugal) and Biblioteca del Real Observatorio de la Marina (San Fernando, Spain).
Support from the Junta de Extremadura and Ministerio de Ciencia e Innovaci\'on of the Spanish Government (AYA2008-04864/AYA) is gratefully acknowledged.
GAK acknowledge visiting grant from the Finnish Academy.


\clearpage



%
\begin{table}
\caption{
Sunspot observations (date, number of sunspot groups $G$ and the observer) for 1636--1642 used
 in this stay.
The last column comments on the relation of the observation to the HS98 \citep{hoyt98} database.
\label{Tab1}}
\begin{tabular}{rccl}
\hline
\hline
Date & $G$ & Observers & Comment\\
\hline
19-20 Jan 1637 & 1 & Marcgraf & New data \\
22 Jan 1637 & 0 & Marcgraf & New data\\
5 Feb 1637 & 1 & Marcgraf & New data\\
9-12 Jun 1637 & 1 & Marcgraf & New data\\
13 Jun 1637 & 0 & Marcgraf & New data\\
21 Sep 1637 & 2 & Marcgraf & New data\\
24-25 Sep 1637 & 2 & Marcgraf & New data\\
28 Sep 1637 & 0 & Marcgraf & New data\\
10 Oct 1637 & 0 & Marcgraf & New data\\
12-13 Oct 1637 & 1 & Marcgraf & New data\\
15 Oct 1637 & 0 & Marcgraf & New data\\
1-3 Jun 1638 & 2 & Horrox & Corrected date \\
29 Oct1638 & 0 & Gassendi & HS98\\
30 Oct-1 Nov 1638 & 1 & Horrox & Corrected date and $G$\\
30 Oct-1 Nov 1638 & 1 & Gassendi & HS98 \\
4 Dec 1639 & 1 & Horrox & New data\\
21-22 Aug 1640 & 1 & Scheiner & HS98 \\
9-22 Jun 1642 & 1 & Rheita & Corrected date and $G$\\
26 Oct 1642 & 0 & Hevelius & HS98\\
28 Oct 1642 & 1 & Hevelius & HS98\\
31 Oct-1 Nov 1642 & 1 & Hevelius & HS98 \\
3-4 Nov 1642 & 1 & Hevelius & HS98\\
6 Nov 1642 & 3 & Hevelius & HS98\\
8 Nov 1642 & 2 & Hevelius & HS98\\
9 and 11-17 Nov 1642 & 1 & Hevelius & HS98\\
18 Nov 1642 & 0 & Hevelius & HS98\\
\hline
\end{tabular}
\end{table}
\begin{table}
\caption{
Yearly group sunspot numbers for 1637--1642: formal values R$_{\rm g}$ \citep{hoyt98} as well as the calculated here
 weighted values R$_{\rm w}$ with $\pm\sigma$ uncertainties.
 \label{Tab2}}
\begin{tabular}{ccc}
\hline
\hline
Year & R$_{\rm g}$ & R$_{\rm w}$  \\
\hline
1637 & n.a. & 13$\pm$2 \\
1638 & 68.7 & 19$\pm$6 \\
1639 & 76.8 & 21$\pm$15 \\
1640 & 15 & 17$\pm$12 \\
1641 & n.a. & n.a. \\
1642 & 47.3 & 13$\pm$3 \\
\hline
\end{tabular}
\end{table}
\clearpage

\clearpage

\end{document}